\documentclass[twocolumn,preprintnumbers,superscriptaddress,nofootinbib,aps,prd,floatfix]{revtex4-2}
\pdfoutput=1
\usepackage{enumerate}
\usepackage{gensymb}
\usepackage{amsmath,amssymb}
\usepackage{mathrsfs}
\usepackage{graphicx}
\usepackage{slashed}
\usepackage{xspace,slashed}
\usepackage{hyperref}
\hypersetup{colorlinks=true, citecolor=blue, urlcolor=blue, linkcolor=blue}
\usepackage[normalem]{ulem}
\usepackage{subfigure,orcidlink}
\usepackage[autostyle]{csquotes}
\usepackage{multirow,array}
\usepackage{float}
\usepackage{hyperref}
\hypersetup{colorlinks=true, citecolor=blue, urlcolor=blue, linkcolor=blue}
\usepackage{tabularray}
\UseTblrLibrary{booktabs}
\usepackage[absolute,overlay]{textpos}

\begin{document}


 \title{Boosted or Inelastic? Discriminating Interpretations of the LZ 248 keV Event}

\author{Satyabrata Mahapatra\,\orcidlink{0009-0009-9592-4786}}
\email{satyabrata@iitgoa.ac.in}
\affiliation{School of Physical Sciences, Indian Institute of Technology Goa, Ponda-403401, Goa, India.}

\author{Partha Kumar Paul\orcidlink{0000-0002-9107-5635}}
\email{ph22resch11012@iith.ac.in}
\affiliation{Department of Physics, Indian Institute of Technology Hyderabad, Kandi, Sangareddy, Telangana-502285, India.}
	
\date{\today}
\begin{abstract}
The LUX-ZEPLIN experiment has reported a single nuclear recoil candidate at $E_R = 248 \pm 23(\mathrm{stat}) \pm 23(\mathrm{sys})$~keV, disfavouring the background-only hypothesis at a global significance of $2.6\sigma$. The difficulty such an event poses is not the recoil energy itself but the absence of any accompanying excess at low energy as elastic scattering of halo dark matter yields a monotonically falling spectrum, and supplying the required momentum transfer $q \simeq 246$~MeV already demands $m_\chi \gtrsim 79$~GeV. We confront the event with the two kinematically distinct mechanisms that evade this limitation, endothermic inelastic dark matter, in which a mass splitting
$\delta \sim \mathcal{O}(100)$~keV forbids low-energy recoils and boosted dark matter, in which a light relativistic flux supplies the momentum, treating both with the same model-independent scalar--scalar, pseudoscalar--scalar and pseudoscalar--pseudoscalar effective operators. The two scenarios prove spectrally distinguishable. The inelastic spectra sit near the observed energy
for every operator, placing only $6$--$17\%$ of events below 150~keV, whereas the
boosted spectra depend critically on the operator: the scalar and pseudoscalar--scalar interactions place $99\%$ and $92\%$ of their events below
150~keV, while the pseudoscalar--pseudoscalar interaction places $75\%$ above it. Momentum dependence is thus essential to the boosted interpretation, but in the inelastic case it trades against the splitting, the preferred $\delta$ decreasing monotonically from $\mathcal{O}_{ss}$ to $\mathcal{O}_{ps}$ to $\mathcal{O}_{pp}$. Because the scenarios differ across the whole high-energy window, a handful of additional events would separate them, placing the question within reach of the full LZ exposure.
\end{abstract}

\maketitle	
\section{Introduction}\label{sec:intro}

The existence of dark matter (DM) is by now established through a
remarkably diverse set of gravitational probes, spanning galactic
rotation curves, the dynamics of galaxy clusters, strong and weak
gravitational lensing, the acoustic peak structure of the cosmic
microwave background and the growth of large scale
structure~\cite{Zwicky:1933gu,Rubin:1970zza,Clowe:2006eq,Planck:2018vyg}. Together
these fix the present abundance to
$\Omega_{\rm DM}h^2 = 0.120 \pm 0.001$~\cite{Planck:2018vyg}, roughly
five times the baryonic energy density. No Standard Model (SM)
particle can play this role, making DM one of the most compelling
pieces of evidence for physics beyond the SM.
Among the many proposals, weakly interacting massive particles (WIMPs)
remain especially attractive where a state with weak-scale mass and
weak-scale annihilation cross section is thermally produced with
approximately the observed abundance, and the same interaction that
sets the relic density generically induces a coupling to nuclei that
is accessible to terrestrial detectors~\cite{Arcadi:2017kky}.

Direct detection experiments have consequently pushed the
spin-independent (SI) WIMP-nucleon cross section down by several
orders of magnitude over the past two decades, with the leading
terrestrial detectors LZ~\cite{LZ:2024zvo}, XENONnT~\cite{XENON:2025vwd} and
PandaX-4T~\cite{PandaX:2024qfu}, now probing $\sigma_{\rm SI}\sim10^{-48}\,{\rm cm}^2$ near
$m_\chi\sim 30$--$50$~GeV and approaching the irreducible neutrino fog. This
sustained absence of a signal motivates a broader view of what a DM
signal could look like.

Against this backdrop, the recent analysis from LZ experiment has revealed a single
nuclear-recoil candidate at $
E_R = 248\pm23_{\rm (stat)}\pm23_{\rm (sys)}\ {\rm keV},
$
in a region with a low expected background, using an exposure of $2.84$ tonne-years and an extended nuclear-recoil energy window reaching approximately $270$ keV \cite{LZ:2026axp}. The background-only hypothesis is disfavored at a global significance of $2.6\sigma$, with a maximum local significance of $3.4\sigma$ across the models considered. The LZ result has already motivated a number of possible DM interpretations \cite{Freese:2026sga,Lou:2026idn,Yin:2026jnn,DiMauro:2026ldr,Visinelli:2026kgt,Yamashita:2026ump,Rodd:2026tyn,McCabe:2026crm,Jeesun:2026vzo,Unwin:2026rdp,Baer:2026fpy,Das:2026uyy,Alhazmi:2026efz,Ahmed:2026qjg,Du:2026lpa,Kannike:2026qyl,Bose:2026ndd,Bandyopadhyay:2026gjw,Borah:2026zwf,Bisal:2026khf,Cheung:2026byg,Elahi:2026vlm,Zhu:2026dag,Lee:2026jxl,Khan:2026nwp,Langhoff:2026ujr,He:2026hqz,Kumar:2026lgi,Heikinheimo:2026kwp}. 
Since the maximum momentum transfer available
in a two-body collision is $q_{\max}=2p^\ast$, with $p^\ast$ the
centre-of-mass momentum, the event demands $p^\ast\gtrsim123$~MeV.
There are only two ways to supply it. Either the DM is heavy, so that
even a halo velocity $v\lesssim v_{\rm esc}+v_e\simeq772$~km/s yields
$p^\ast=\mu_{\chi N}v\gtrsim123$~MeV, which requires
$\mu_{\chi N}\gtrsim48$~GeV and hence $m_\chi\gtrsim79$~GeV; or the DM
is light but {relativistic}.

One particularly attractive possibility is inelastic dark matter (iDM), in which the incident DM particle scatters into a state with a slightly different mass~\cite{Tucker-Smith:2001myb,Tucker-Smith:2004mxa,Cui:2009xq, Borah:2020smw, Cho:2024lhp}. If the dark sector
contains two nearly degenerate states $\chi_1$ and $\chi_2$ split by
$\delta=m_{\chi_2}-m_{\chi_1}$ and coupled off-diagonally to nucleons,
scattering proceeds only through $\chi_1N\to\chi_2N$, and the minimum
velocity required to deposit $E_R$ becomes
$v_{\min}=(m_NE_R/\mu_{\chi N}+\delta)/\sqrt{2m_NE_R}$. For endothermic
scattering with $\delta$ of order $100$~keV, this function is large at
small $E_R$, so low energy recoils are kinematically forbidden
outright; the surviving rate is confined to a band at high recoil
energy whose position is set by $\delta$ and $m_{\chi_1}$. The absence of a
low energy excess thus motivates iDM scenarios.

An alternative possibility is boosted dark matter (bDM), in which the incident DM particles have velocities substantially larger than those expected for virialized Galactic halo DM. Such a population can arise naturally in multicomponent dark sectors, for example through annihilation or decay of a heavier dark-sector state. In this case, the energy of the incident particle is determined primarily by the dark-sector mass spectrum rather than by the Galactic velocity distribution. A light boosted particle can therefore produce a high-energy nuclear recoil even when its mass is well below that required for conventional halo DM to generate the same recoil energy. Such bDM can provide an alternative explanation of the LZ event, either through momentum-dependent interactions or through near-threshold inelastic scattering of an approximately monochromatic boosted flux. We consider bDM~\cite{Agashe:2014yua,Berger:2014sqa,Kim:2016zjx,Fornal:2020npv,Borah:2021jzu}, in which a
sub-dominant light species $\chi_1$ is produced relativistically today
by the annihilation of a heavier component,
$\chi_2\chi_2\to\chi_1\chi_1$, in the Galactic halo. The outgoing
$\chi_1$ carries a Lorentz factor $\gamma=m_{\chi_2}/m_{\chi_1}$ fixed
by the mass ratio, and the resulting monoenergetic flux can transfer
$q\sim\mathcal{O}(100)$~MeV even for $m_{\chi_1}\ll1$~GeV. The recoil
spectrum is then a plateau extending to a sharp kinematic endpoint
$E_R^{\max}$ rather than a falling exponential, again concentrating
events at high energy.

Rather than committing to a specific ultraviolet completion, we work
throughout with a model independent effective operator description, in
the spirit of the effective field theory of
direct detection. Here it is worth mentioning that such an analysis is well motivated, as momentum suppressed
operators produce spectra that {rise} with $E_R$, reinforcing the
high energy concentration already supplied by the inelastic or boosted
kinematics. We, therefore, analyse scalar--scalar, pseudoscalar--scalar
and pseudoscalar--pseudoscalar contact operators on an equal footing
for both the iDM and bDM scenarios.

For each operator and each scenario we construct the full predicted
spectrum, including nuclear form factors, LZ detection efficiency and
energy resolution, and perform a profile likelihood analysis of the
single observed event. For iDM the parameter space is
$(m_{\chi_1},\delta,\Lambda)$; for bDM it is
$(m_{\chi_1},\gamma\ {\rm or}\ v,\Lambda)$, with the boost fixed by
$m_{\chi_2}/m_{\chi_1}$ and the flux normalization controlled by
$\langle\sigma v\rangle_{\chi_2\chi_2\to\chi_1\chi_1}$. {We present the confidence regions in the relevant parameter space and investigate the dependence of the preferred regions on the different effective operators, thereby identifying the operators that provide the best explanation of the observed event.}

\begin{figure}[h]
\centering
\includegraphics[width=1\linewidth]{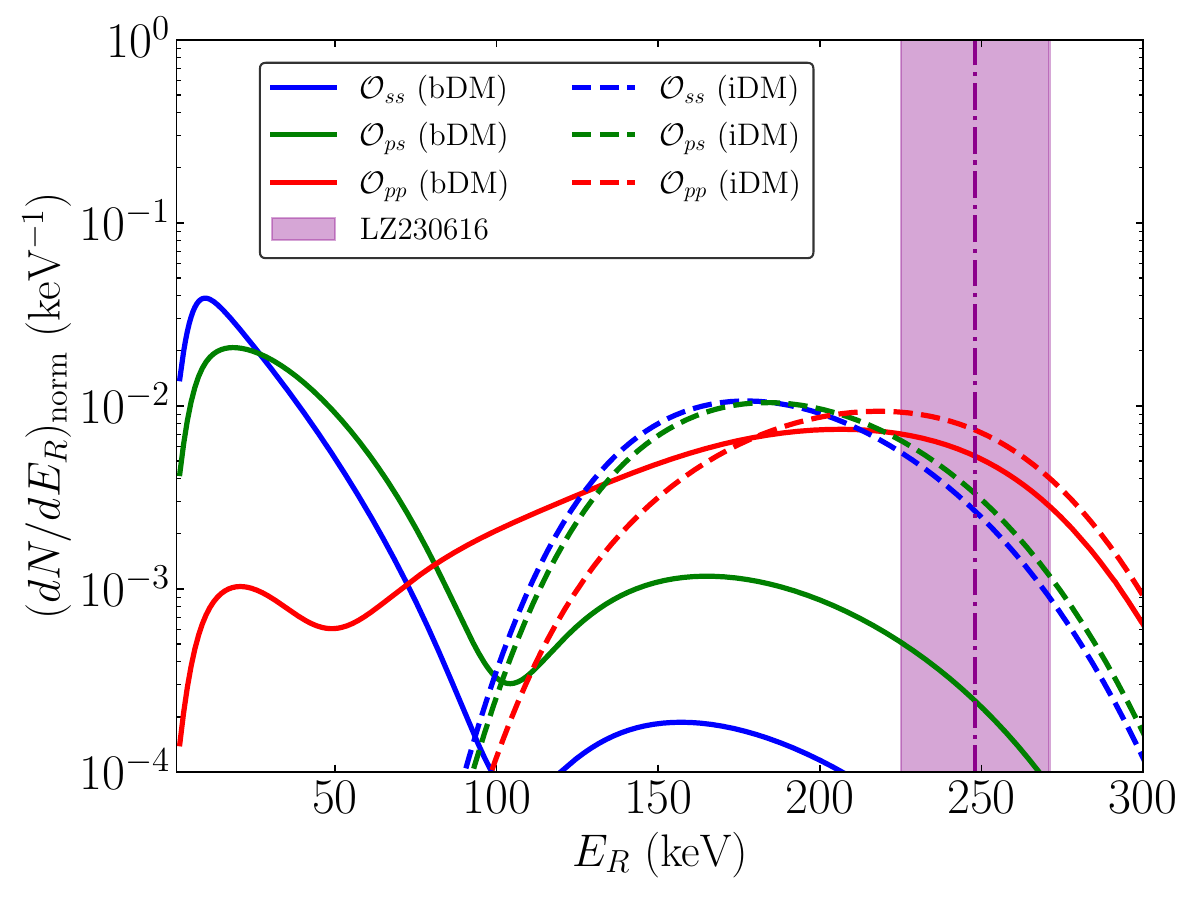}
\caption{Normalized recoil spectra as a function of recoil energy for 6 operators in boosted DM (bDM) and inelastic DM (iDM) scenario are shown. The dark magenta shaded region indicates the recoil energy range of the LZ event ($248{\rm ~keV}\pm23\rm ~keV$), including its uncertainty, while the dash-dotted line marks the central value.}
\label{fig:recoilspectra}
\end{figure}

\section{Effective DM--nucleon interactions}\label{sec:eft}
We describe the DM--nucleon interaction with dimension-six contact
operators suppressed by a cutoff scale $\Lambda$
\begin{align}
\mathcal{O}_{ss}&=\frac{1}{\Lambda^2}(\overline{\chi_1}\chi_2)(\bar NN)\,,
\label{eq:Oss}\\
\mathcal{O}_{ps}&=\frac{1}{\Lambda^2}(\overline{\chi_1}i\gamma_5\chi_2)(\bar NN)\,,
\label{eq:Ops}\\
\mathcal{O}_{pp}&=\frac{1}{\Lambda^2}(\overline{\chi_1}i\gamma_5\chi_2)(\bar Ni\gamma_5N)\,,
\label{eq:Opp}
\end{align}

The distinction among Eqs.~\eqref{eq:Oss}--\eqref{eq:Opp} lies in their
non-relativistic reduction. Following the standard
dictionary~\cite{Fan:2010gt,Fitzpatrick:2012ix,Dent:2015zpa},
$\mathcal{O}_{ss}$ maps onto the operator $\mathbf{1}$ and is
therefore unsuppressed and {coherent}, receiving the full $A^2$
enhancement through the nuclear response. The operator
$\mathcal{O}_{ps}$ carries a pseudoscalar current on the DM side and
the identity on the nucleon side; it is suppressed by
$q^2/4m_{\chi_1}^2$ but retains coherence, so the $A^2F^2(q)$ weighting
survives. The operator $\mathcal{O}_{pp}$ is pseudoscalar on both
sides, giving a $q^4/(16\,m_{\chi_1}^2m_n^2)$ suppression together with
a spin-dependent nuclear response. Thus, only the odd isotopes $^{129}$Xe and
$^{131}$Xe contribute, through the longitudinal spin structure function
$S_L(q)$ \cite{Jeong:2021bpl}.

The corresponding differential cross sections, valid for arbitrary
incident velocity, are:
\begin{eqnarray}
\frac{d\sigma^{ss}}{dE_R}=\frac{m_NA^2F^2(q)}{\pi \Lambda^4}\frac{1-v^2}{v^2}
\label{eq:dsigss}
\end{eqnarray}
\begin{eqnarray}
\frac{d\sigma^{ps}}{dE_R}=\frac{q^2m_NA^2F^2(q)}{4\pi \Lambda^4m_{\chi_1}^2}\frac{1-v^2}{v^2}
\label{eq:dsigps}
\end{eqnarray}
\begin{eqnarray}
\frac{d\sigma^{pp}}{dE_R}=\frac{q^4m_NS_L(q)}{32\pi \Lambda^4m_n^2m_{\chi_1}^2}\frac{1-v^2}{v^2}
\label{eq:dsigpp}
\end{eqnarray}
where $q=\sqrt{2m_NE_R}$ and $v$ is the DM velocity in the detector
frame. Here the kinematic factor $\frac{1-v^2}{v^2}$ can be interpreted as:
\begin{equation}
\frac{1-v^2}{v^2}=\frac{1}{\gamma^2v^2}=\frac{m_{\chi_1}^2}{p_{\chi_1}^2}\,,
\label{eq:kinfactor}
\end{equation}
i.e.\ the cross section scales as the inverse square of the incident
momentum. In the non-relativistic limit $p_{\chi_1}\to m_{\chi_1}v$
this reduces to the familiar $1/v^2$ of halo DM scattering while for a relativistic boosted flux it falls as
$1/\gamma^{2}$, so
Eqs.~\eqref{eq:dsigss}--\eqref{eq:dsigpp} apply uniformly to both the
iDM and bDM analyses.

The momentum dependence of these interactions is important for interpreting a high-energy event. For the $\mathcal{O}_{ss}$ interaction, the recoil spectrum is predominantly controlled by the kinematics and the nuclear form factor. By contrast, the additional powers of $q$ in the pseudoscalar interactions shift relative weight towards larger recoil energies. This provides a simple effective description of the mechanism in which momentum-dependent interactions can reduce the accompanying low-energy recoil events. {In Fig. \ref{fig:recoilspectra}, we show the normalized recoil spectra for the three operators in two cases: iDM and bDM. The details of which are discussed in Sec. \ref{sec:lz}.}

\section{Endothermic inelastic dark matter}\label{sec:idm}

For a two-state dark sector with splitting
$\delta=m_{\chi_2}-m_{\chi_1}>0$, energy-momentum conservation in
$\chi_1N\to\chi_2N$ requires a minimum incident speed
\begin{equation}
v_{\min}(E_R)=\frac{1}{\sqrt{2m_NE_R}}
\left|\frac{m_NE_R}{\mu_{\chi N}}+\delta\right|\,,
\label{eq:vmin}
\end{equation}
with $\mu_{\chi N}=m_{\chi_1}m_N/(m_{\chi_1}+m_N)$. The exothermic case
corresponds to $\delta\to-\delta$. Eq.~\eqref{eq:vmin} is the origin of
the spectral shape that makes inelastic DM (iDM) relevant here.
Unlike the elastic case, in which $v_{\min}$ decreases monotonically as
$E_R\to0$, the presence of $\delta$ makes the second term dominate at
small recoil energy, so that $v_{\min}$ diverges as $E_R^{-1/2}$ and the
low-energy rate is switched off entirely. The function attains its
minimum at
\begin{equation}
E_R^\star=\frac{\mu_{\chi N}\,\delta}{m_N}\,,
\qquad
v_{\min}(E_R^\star)=\sqrt{\frac{2\delta}{\mu_{\chi N}}}\equiv v_{\rm thr}\,,
\label{eq:ERstar}
\end{equation}
the latter being simply the statement that the centre-of-mass kinetic
energy $\tfrac12\mu_{\chi N}v^2$ must supply the mass splitting.
Inverting Eq.~\eqref{eq:vmin} at fixed speed gives the accessible
window
\begin{equation}
E_R^{\pm}(v)=\frac{\mu_{\chi N}^2}{2m_N}
\left(v\pm\sqrt{v^2-\frac{2\delta}{\mu_{\chi N}}}\right)^{\!2}\,,
\label{eq:ERband}
\end{equation}
which satisfies $E_R^+E_R^-=(E_R^\star)^2$ and reduces to
$E_R^-=0$, $E_R^+=2\mu_{\chi N}^2v^2/m_N$ in the elastic limit. The
observable signal is therefore confined to a band in recoil energy
rather than following a falling exponential. For the benchmark
$m_{\chi_1}=1$~TeV and $\delta=278$~keV, chosen such that
$E_R^\star=248$~keV, one finds $v_{\rm thr}=678$~km/s and, at
$v=v_{\max}$, we get window $87.5~{\rm keV}\le E_R\le702$~keV {\it i.e.} no recoil
below $87.5$~keV is kinematically permitted at any halo velocity. The
absence of a low-energy excess accompanying the $248$~keV event is thus
a direct consequence of the inelastic kinematics rather than a
suppression imposed by hand.

Demanding $v_{\min}(E_R)\le v_{\max}=v_{\rm esc}+v_e$ bounds the
accessible parameter space independently of the operator choice. With
$v_{\rm esc}=540$~km/s~\cite{Smith:2006ym,Deason:2019tpl} and $v_e=232$~km/s, the
splitting is limited by
\begin{equation}
\delta\le v_{\max}\sqrt{2m_NE_R}-\frac{m_NE_R}{\mu_{\chi N}}\,,
\label{eq:deltamax}
\end{equation}
which vanishes at $m_{\chi_1}\simeq79$~GeV, reaches
$\delta_{\max}\simeq325$~keV at $m_{\chi_1}=500$~GeV, and asymptotes to
$\delta_{\max}\simeq385$~keV for $m_{\chi_1}\gg m_N$. We overlay this
boundary on all parameter space figures and it is a hard kinematic edge
that no choice of cross section can evade.

The differential event rate can be evaluated by convolving
Eqs.~\eqref{eq:dsigss}--\eqref{eq:dsigpp} with the local velocity
distribution,
\begin{equation}
\frac{d\mathcal{R}}{dE_R}=N_T\frac{\rho_{\chi_1}}{m_{\chi_1}}
\int_{v_{\min}(E_R)}^{v_{\max}}dv\,v\,f(v)\,\frac{d\sigma}{dE_R}\,,
\label{eq:rateiDM}
\end{equation}
where $N_T$ is the number of target nuclei per unit detector mass and
$\rho_{\chi_1}\simeq0.4$~GeV\,cm$^{-3}$ is the local DM density. Here $f(v)$ is given by
\begin{eqnarray}
    f(v)=\frac{v}{\sqrt{\pi}v_ev_0}e^{-\frac{v_e^2+v^2}{v_0^2}}\left( e^{\frac{2vv_e}{v_0^2}}- e^{-\frac{2vv_e}{v_0^2}} \right),
\end{eqnarray}
with $v_0=220$ km/s, and the time-averaged Earth's speed relative to the galactic rest frame is $v_e=v_\odot=v_0+12 {~\rm km/s}$. In the numerical analysis we adopt this Standard Halo Model (SHM) as the reference velocity distribution. The important point, however, is that the endothermic interpretation is inherently sensitive to the high-velocity tail because the required splitting is close to the kinematic threshold.

The three parameters relevant for the iDM analysis are consequently
\begin{equation}
\{m_\chi, \delta, \Lambda\}.
\end{equation}
For each point in this parameter space we calculate the recoil spectrum, convolve it with the LZ detector efficiency and energy resolution, and compare the resulting spectral shape with the observed event.

\section{Boosted dark matter}\label{sec:bdm}

We next consider a qualitatively different origin for the high-energy nuclear recoil. Instead of relying on the high-velocity tail of the Galactic DM distribution, we assume that a population of DM particles arrives at the detector with a velocity substantially larger than the virial velocity of halo DM. Such particles are commonly referred to as boosted DM (bDM). A simple realization is a multicomponent dark sector containing a heavier state $\chi_2$ and a lighter state $\chi_1$ with $\chi_2$ constituting the bulk of the relic abundance and annihilates in
the Galactic halo into the lighter state,
$\chi_2\chi_2\to\chi_1\chi_1$~\cite{Agashe:2014yua,Berger:2014sqa}.  Since $\chi_2$ is non-relativistic today, each outgoing $\chi_1$
carries energy $E_{\chi_1}\simeq m_{\chi_2}$, so the flux is monoenergetic with the differential energy spectrum 
\begin{equation}
    \frac{dN_{\chi_1}}{dE_{\chi_1}}=2 \delta(E_{\chi_1}-m_{\chi_2})\,.
\end{equation}
and 
\begin{eqnarray}
\gamma=\frac{E_{\chi_1}}{m_{\chi_1}}=\frac{m_{\chi_2}}{m_{\chi_1}}\,,\nonumber\\
v_{\chi_1}=\sqrt{1-\gamma^{-2}}\,,
\nonumber\\
p_{\chi_1}=m_{\chi_1}\sqrt{\gamma^2-1}\,,
\label{eq:boost}
\end{eqnarray}
Here, the important distinction from ordinary halo DM is that the incident kinetic energy of the boosted dark particle is now an independent physical scale. The boost is fixed by the
mass ratio of the two dark states.

The resulting monochromatic flux of $\chi_1$ at Earth is given by
\begin{equation}
\Phi^{\rm GC}_{\chi_2}=\frac{\langle\sigma v\rangle_{\chi_2\chi_2\to\chi_1\chi_1}}
{4\pi m_{\chi_2}^2}\int d\Omega\int_{\rm l.o.s.}\!\!d\ell\,\rho_{\chi_2}^2\,,
\label{eq:fluxgeneral}
\end{equation}
which for an NFW profile with $\rho_\odot=0.4$~GeV\,cm$^{-3}$ and
$R_\odot=8.5$~kpc \cite{Navarro:1995iw,Genolini:2021doh}, integrated over the full sky, evaluates to \cite{Agashe:2014yua,Borah:2021yek}
\begin{equation}
\Phi^{\rm GC}_{\chi_2}=6\times10^{3}\,{\rm cm^{-2}\,s^{-1}}
\left(\frac{\langle\sigma v\rangle_{\chi_2\chi_2\to\chi_1\chi_1}}
{3.5\times10^{-31}\,{\rm cm^{2}}}\right)
\left(\frac{0.1\,{\rm GeV}}{m_{\chi_2}}\right)^{2}.
\label{eq:fluxnum}
\end{equation}

This boosted flux of Eq.~\eqref{eq:fluxnum} scales linearly with
$\langle\sigma v\rangle_{\chi_2\chi_2\to\chi_1\chi_1}$ and inversely
with the square of the mass of the annihilating state $\chi_2$.
Reproducing the observed event rate therefore requires an annihilation
cross section larger than the canonical thermal value,
and it is necessary to verify that such a rate does not deplete the
$\chi_2$ population over the lifetime of the Galaxy, nor remove an
appreciable fraction of the halo mass through the escape of the
relativistic $\chi_1$ produced.

Within a virialized halo the cosmological expansion term is negligible,
and the number density of the annihilating component obeys
\begin{eqnarray}
\frac{dn_{\chi_2}}{dt}=-\Gamma(\chi_2\chi_2\to\chi_1\chi_1)\,n_{\chi_2}\,,
\nonumber\\{\rm with ~~}
\Gamma=n_{\chi_2}\langle\sigma v\rangle_{\chi_2\chi_2\to\chi_1\chi_1}\,.
\label{eq:boltzB}
\end{eqnarray}
Separating variables and
integrating from the epoch of galaxy formation to the present gives
\begin{equation}
\frac{1}{n^{\rm today}_{\chi_2}}=\frac{1}{n^{\rm init}_{\chi_2}}
+\langle\sigma v\rangle_{\chi_2\chi_2\to\chi_1\chi_1}\,t_{\rm MW}\,,
\label{eq:nsol}
\end{equation}
or equivalently
\begin{equation}
n^{\rm today}_{\chi_2}=\frac{n^{\rm init}_{\chi_2}}{1+\mathcal{D}}\,,
\qquad
\mathcal{D}\equiv n^{\rm init}_{\chi_2}\,
\langle\sigma v\rangle_{\chi_2\chi_2\to\chi_1\chi_1}\,t_{\rm MW}\,.
\label{eq:depletion}
\end{equation}
Here, the dimensionless
parameter $\mathcal{D}$ measures the number of annihilations experienced
by a typical $\chi_2$ particle over the lifetime of the Galaxy and
$t_{\rm MW}=13.61$~Gyr ($=4.30\times10^{17}$~s $=1.29\times10^{28}$~cm)
is the age of the Milky Way and $n^{\rm init}_{\chi_2}$ is the number
density at the epoch of its formation. The abundance is therefore unchanged provided $\mathcal{D}\ll1$. For the
cross sections required to explain the LZ event, $\mathcal{D}\lesssim10^{-2}$, so
that both the $\chi_2$ abundance and the total halo mass are preserved
to better than one percent over the age of the Galaxy. The scenario is
thus self-consistent as the annihilation responsible for the boosted flux
neither exhausts its own source nor causes appreciable evaporation of
the DM halo.

Because the incident $\chi_1$ is relativistic, the recoil kinematics
must be treated exactly. For a particle of mass $m_{\chi_1}$, energy
$E_{\chi_1}=\gamma m_{\chi_1}$ and momentum $p_{\chi_1}$ scattering
elastically off a nucleus of mass $m_N$, the maximum recoil energy is
\begin{equation}
E_R^{\max}=\frac{2m_Np_{\chi_1}^2}
{(m_N+m_{\chi_1})^2+2m_N(E_{\chi_1}-m_{\chi_1})}\,,
\label{eq:ERmax}
\end{equation}
the denominator being the Mandelstam invariant $s$. For
$m_{\chi_1}\ll m_N$ this reduces to $E_R^{\max}\simeq2p_{\chi_1}^2/m_N$,
so that reaching $E_R=248$~keV requires
$p_{\chi_1}\gtrsim\sqrt{m_NE_R/2}\simeq123$~MeV, independently of
$m_{\chi_1}$.

It is useful to keep the production and scattering parameters conceptually separate. The scattering spectrum is determined by $(m_\chi,v_\chi,\Lambda)$ and the choice of effective operator, while the overall number of events is proportional to the incident bDM flux. If the boosted population is generated by annihilation of a heavier DM component, the flux normalization can subsequently be translated into a constraint on the annihilation cross section.

The parameter space relevant for the bDM interpretation can therefore be represented by
\begin{equation}
\{
m_\chi\,,
\Lambda\,,
v_\chi\,,
\Phi_\chi
\},
\end{equation}
or equivalently by
\begin{equation}
\{
m_\chi\,,
\Lambda\,,
v_\chi\,,
\langle\sigma v\rangle_{\chi_2\chi_2\to\chi_1\chi_1}
\}
\end{equation}
once a specific production mechanism is assumed.

The principal phenomenological advantage of bDM is that the incident energy is not tied to the high-velocity tail of the Galactic halo with higher uncertainty. This provides a qualitatively different explanation of the LZ event from endothermic halo iDM.

\section{LZ Event Analysis}\label{sec:lz}

{In this section, we study the compatibility of both the iDM and bDM scenarios with the observed LZ230616 event. We calculate the expected number of signal events for the relevant effective operators, taking into account the nuclear form factor, detector efficiency and recoil-energy smearing. We then perform a likelihood analysis to identify the preferred regions of the parameter space that can accommodate the observed event.} The differential number of signal events per unit of true recoil energy is
\begin{equation}
\frac{dN_s^{i}}{dE_R}=\Phi_\chi \frac{\mathcal{E}_{\rm LZ}}{m_N}\epsilon(E_R)\frac{d\sigma^i}{dE_R},\label{eq:dNdeR_bdm}
\end{equation}
with $i\in\{ss,ps,pp\}$, where $\mathcal{E}_{\rm LZ}=2.84$~tonne-years
is the exposure and $\epsilon(E_R)$ is the nuclear recoil detection
efficiency \cite{LZ:2026axp}.
Helm form factor is
\begin{eqnarray}
F^2(E_R)=\left( \frac{3j_1(qr_0)}{qr_0} \right)^2e^{-s^2q^2},
\end{eqnarray}
where $q=\sqrt{2m_NE_R}$, $s=1$ fm, $r_0=\sqrt{r^2-5s^2}$, $r=1.2A^{1/3}$ fm. With smearing for the $\mathcal{O}_{ss}$ operator
\begin{align}
\frac{dN^{ss}_s}{dE_{\rm obs}}=\Phi_\chi \frac{\mathcal{E}_{\rm LZ}}{m_N} \frac{m_NA^2}{\pi\Lambda^4}\frac{1-v^2}{v^2}\int_0^{E_R^{\rm max}} dE_R F^2(E_R)\epsilon(E_R)\mathcal{G}_{\rm sm}
\label{eq:smear}
\end{align}
where the smearing factor is given as
\begin{eqnarray}
\mathcal{G}_{\rm sm}=\frac{1}{\sqrt{2\pi}\sigma_E}e^{-\frac{(E_{\rm obs}-E_R)^2}{2\sigma_E^2}}
\end{eqnarray}
with $\sigma_E=1.46{~\rm keV}\sqrt{E_R/{\rm keV}}$.
The total number of expected events in the analysis window
$[E_l,E_h]=[5.4,270]$~keV is obtained by integrating
Eq.~\eqref{eq:smear} over $E_{\rm obs}$,
\begin{equation}
N^{ss}_s=\int_{E_l}^{E_h}dE_{\rm obs}\frac{dN^{ss}_s}{dE_{\rm obs}}\,.
\end{equation}
Interchanging the order of
integration allows the observed-energy integral to be performed
analytically, giving
\begin{equation}
N^{ss}_s=\Phi_\chi\,\mathcal{E}_{\rm LZ}\frac{A^2}{\pi \Lambda^4}
\frac{1-v^2}{v^2}\int_0^{E_R^{\max}}\!\!dE_R\,F^2(E_R)\,
\epsilon(E_R)\,\eta(E_R)\,,
\label{eq:Ns}
\end{equation}
where the window acceptance function is
\begin{equation}
\eta(E_R)=\frac{1}{2}\left[
{\rm erf}\!\left(\frac{E_h-E_R}{\sqrt2\,\sigma_E}\right)
-{\rm erf}\!\left(\frac{E_l-E_R}{\sqrt2\,\sigma_E}\right)\right].
\label{eq:eta}
\end{equation}

Equation~\eqref{eq:Ns} is written for the scalar--scalar case. The
pseudoscalar operators follow by inserting the corresponding
$q$-dependent factors and, for $\mathcal{O}_{pp}$, replacing
$A^2F^2(E_R)$ by $S_L(q)$. {The above expressions are for the bDM case. The iDM case can be easily recovered by replacing $\Phi_\chi\frac{d\sigma^i}{dE_R}$ in Eq. (\ref{eq:dNdeR_bdm}) with $\frac{\rho_{\chi_1}}{m_{\chi_1}}\int dvvf(v)\frac{d\sigma^i}{dE_R}$. Here $f(v)$ is the DM speed distribution.}

With a single observed event and negligible background in the high
energy region, we use an extended unbinned likelihood. Defining the
normalized signal spectrum
\begin{equation}
f^i_s(E_{\rm obs})=\frac{1}{N^i_s}\frac{dN^i_s}{dE_{\rm obs}}
\label{eq:fs}
\end{equation}
the likelihood for $n_0$ observed events is
\begin{equation}
\mathcal{L}=e^{-N^i_s}\prod_{i=1}^{n_0}N^i_s\,f^i_s(E_i)\,,
\label{eq:like}
\end{equation}
which for $n_0=1$ gives
$\ln\mathcal{L}=-N^i_s+\ln N^i_s+\ln f^i_s(E_{\rm obs})$.

Minimizing w.r.t $\Lambda$, we get
\begin{equation}
\ln\mathcal{L}_{\rm prof}=-1+\ln f^i_s(E_{\rm obs},m_{\chi_1},\theta))
\end{equation}
where $\theta$ denotes $\delta$ for iDM and $v$ (or equivalently $\gamma$)
for bDM. The scale $\Lambda$ thus drops out of the shape comparison entirely, and the test statistic
\begin{equation}
TS=\Delta\chi^2=-2\ln\left( \frac{f^i_s(E_{\rm obs},m_{\chi_1},v({\rm or~}{\delta}))}{f^i_s(E_{\rm obs},\hat{m}_{\chi_1},\hat{v}({\rm or~}{\delta})) } \right)
\end{equation}
measures only how well each point in parameter space places the
predicted spectrum at the observed recoil energy. Confidence region
is drawn at $\Delta\chi^2=2.30$ for
$1\sigma$ in two parameters.

\begin{figure*}[t]
\centering
\includegraphics[width=1\linewidth]{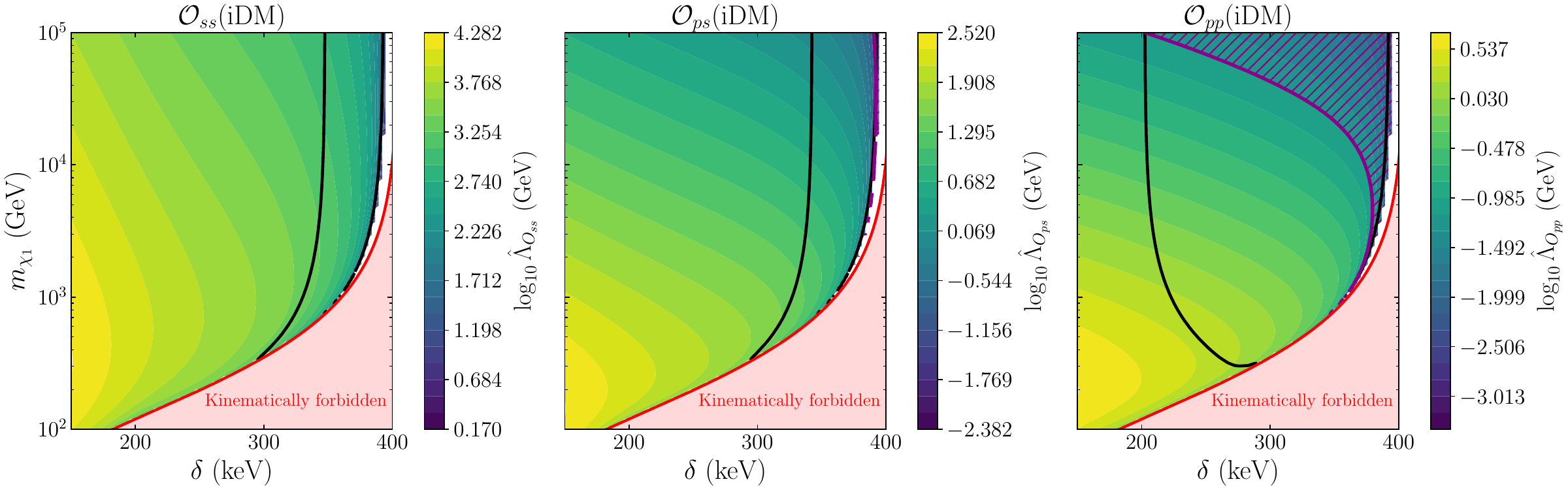}
\caption{Profiled interaction scale $\hat{\Lambda}_{\mathcal{O}_{i}}$ in the $(m_{\chi_1},\delta)$ plane for the iDM scenario $\left(\mathcal{O}_{ss}[\rm{\textit{left}}],\mathcal{O}_{ps}[\rm{\textit{middle}}],\mathcal{O}_{pp}[\rm{\textit{right}}]\right)$. At each point, $\Lambda$ is optimized to obtain the best fit to the observed $248~\mathrm{keV}$ recoil, and the corresponding $\hat{\Lambda}_{O_{i}}$ is shown by the color code. The red solid curve represents the kinematic boundary $\delta_{\max}=\frac12\mu_Nv_{\max}^2$, above which the inelastic transition is inaccessible for the maximum DM velocity $v_{\max}$. The black solid curve corresponds to $TS=2.3$, defining the $1\sigma$ confidence region in the $(m_{\chi_1},\delta)$ plane. {The dark magenta hatched region corresponds to $\hat{\Lambda}<69.4$ MeV. See main text for details.}}
\label{fig:idm_prof}
\end{figure*}

To illustrate the spectral differences between the two DM scenarios, we show the normalized recoil-energy spectra for the three operators considered in the bDM and iDM scenarios in Fig. \ref{fig:recoilspectra}. The solid curves denote the bDM operators whereas dashed colored curves denotes the iDM operators. The iDM spectra are mostly concentrated toward the high-recoil region relevant for the 248 keV event. On the other hand, the bDM spectra shows a substantially stronger operator dependence and, contain large low-energy components. For $\mathcal{O}_{ss}$ and $\mathcal{O}_{ps}$, the bDM spectra contain a larger fraction of events at low recoil energies, while the corresponding iDM spectra are shifted toward higher recoil energies. The $\mathcal{O}_{pp}$ bDM spectrum has a larger fraction of events at high recoil energies and a more pronounced high-energy tail. To quantify these differences independently of the overall normalization, we define the spectral fraction
\begin{eqnarray}
\mathcal{SF}_{150}=\frac{N_s(E_R<150{~\rm keV})}{N_s^{\rm tot}},
\end{eqnarray}
where $N_s(E_R<150{~\rm keV})$ is the number of signal event with energy $E_R<150$ keV, and $N_s^{\rm tot}$ is the total number of event. A larger $\mathcal{SF}_{150}$ means that a larger fraction of events occur below 150 keV, while a smaller \(\mathcal{SF}_{150}\) means that more events occur at higher recoil energies. For illustration purpose, we fix $m_{\chi_1}=1~\mathrm{TeV}$ and $\delta=300~\mathrm{keV}$. Since the overall normalization cancels in this ratio, $\mathcal{SF}_{150}$ is independent of the interaction scale $\Lambda$. We obtain $\mathcal{SF}_{150}=17.44\%$, $14.11\%$, and $6\%$ for the $\mathcal{O}_{ss}$, $\mathcal{O}_{ps}$, and $\mathcal{O}_{pp}$ operators, respectively. Thus, the $\mathcal{O}_{ss}$ spectrum has the largest fraction of low-energy events, followed by $\mathcal{O}_{ps}$, while $\mathcal{O}_{pp}$ has the smallest fraction. In other words, $\mathcal{O}_{pp}$ places a larger fraction of its events at recoil energies above $150~\mathrm{keV}$. This shows that the three operators predict noticeably different recoil-energy distributions even for the same DM mass and mass splitting.

Now moving to the bDM scenario, the boosted nature of the incoming DM can lead to substantially different recoil-energy distributions for different operators. To illustrate this, we fix $m_{\chi_1}=10~\mathrm{GeV}$ and $p_\chi=500~\mathrm{MeV}$, corresponding to $v\simeq0.05c$. For the $\mathcal{O}_{ss}$, $\mathcal{O}_{ps}$, and $\mathcal{O}_{pp}$ operators, we obtain $\mathcal{SF}_{150}=99\%$, $92\%$, and $25\%$, respectively. {Thus, the $\mathcal{O}_{ss}$ and $\mathcal{O}_{ps}$ operators predict that most of the events should occur below $150~\mathrm{keV}$, whereas the $\mathcal{O}_{pp}$ operator places about $75\%$ of the events above $150~\mathrm{keV}$, in particular $22\%$ of the total events in the LZ230616 recoil window (225 keV to 271 keV). On the other hand, $\mathcal{O}_{ss}$ and $\mathcal{O}_{ps}$ place only $0.14\%$ and $1.2\%$ events, respectively.} This difference is particularly relevant for the $248~\mathrm{keV}$ recoil observed by LZ. If the interaction were dominated by $\mathcal{O}_{ss}$ or $\mathcal{O}_{ps}$, a large fraction of the predicted events would be expected at lower recoil energies. In contrast, $\mathcal{O}_{pp}$ naturally gives a much larger fraction of events in the high-recoil region containing the observed event. Therefore, the recoil-energy distribution of the LZ event provides a preference for the $\mathcal{O}_{pp}$ operator over $\mathcal{O}_{ss}$ and $\mathcal{O}_{ps}$ for this bDM benchmark.

The left panel of Fig. \ref{fig:idm_prof} shows the profile-likelihood result for the $\mathcal{O}_{ss}$ operator in the $\{m_{\chi_1},\delta\}$ plane. At each point, the interaction scale $\Lambda$ is varied and optimized, and the corresponding best fit value $\hat{\Lambda}_{\mathcal{O}_{ss}}$ is shown by the color scale. The red solid curve represents the kinematic boundary $\delta_{\max}=\frac12\mu_{\chi N}v_{\max}^2$, above which the inelastic transition is inaccessible for the maximum DM velocity $v_{\max}$. Thus, the region below it is kinematically forbidden. The resulting boundary exhibits a clear correlation between $m_{\chi_1}$ and $\delta$. As the mass splitting is increased, a larger DM mass is required to provide sufficient energy for the inelastic transition, which can result in the observed nuclear recoil. The solid black curve shows the $TS=2.3$ contour, corresponding to the $1\sigma$ region in the two dimensional parameter space after profiling over $\Lambda$. Thus, while the red curve determines where the observed recoil is kinematically possible, the black contour identifies the region that provides a statistically preferred description of the event. The color variation within the allowed region further shows how the interaction scale required to describe the event changes across the $m_{\chi_1}-\delta$ parameter space.

\begin{figure*}[t]
\centering
\includegraphics[width=1\linewidth]{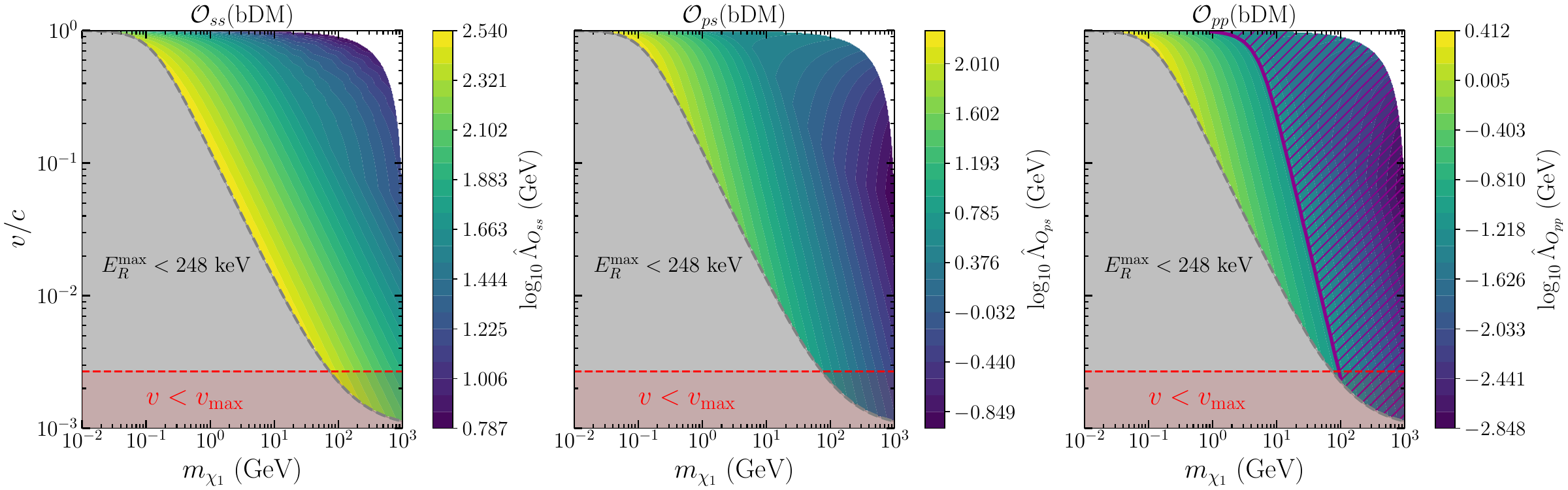}
\caption{Velocity as a function of DM mass for the three operators in the bDM case $\left(\mathcal{O}_{ss}[\rm{\textit{left}}],\mathcal{O}_{ps}[\rm{\textit{middle}}],\mathcal{O}_{pp}[\rm{\textit{right}}]\right)$ is shown with different colored lines. The black dashed line with the gray shaded region represents $E_{R}^{\rm max}=248$ keV. In the red shaded region $v<v_{\rm max}$, thus no boost is required in this region. The dark magenta hatched region corresponds to $\hat{\Lambda}<69.4$ MeV. See main text for details.}
\label{fig:vVSmchi}
\end{figure*}

The middle panel of Fig. \ref{fig:idm_prof} shows the result for the $\mathcal{O}_{ps}$ operator. The kinematic boundary remains the same as in the $\mathcal{O}_{ss}$ case, since it is determined only by the maximum DM velocity. However, the $TS=2.3$ contour is shifted slightly towards smaller $\delta$ for given $m_{\chi_1}$. This shift can be understood from the recoil energy dependence of the $\mathcal{O}_{ps}$ operator. Compared with $\mathcal{O}_{ss}$, the $\mathcal{O}_{ps}$ operator contains an additional $q^2$ factor. Since $q^2$ increases with $E_R$, this gives relatively more weight to larger recoil energies. Therefore, even for the same $m_{\chi_1}$ and $\delta$, the $\mathcal{O}_{ps}$ operator produces relatively more events at higher recoil energies than $\mathcal{O}_{ss}$. The minimum velocity required for a recoil contains a contribution proportional to $\delta/\sqrt{E_R}$ (See Eq. \ref{eq:vmin}). This means that the effect of increasing $\delta$ is larger for smaller recoil energies. For a low energy recoil, $1/\sqrt{E_R}$ is larger, so the increase in $v_{\rm min}$ caused by a given increase in $\delta$ is larger. As a result, increasing $\delta$ suppresses low energy recoils more strongly, while the higher energy part of the spectrum is less affected. In this way, increasing $\delta$ also increases the relative fraction of events at higher recoil energies.
Thus, the $q^2$ dependence of $\mathcal{O}_{ps}$ and the effect of increasing $\delta$ act in the same direction. The $\mathcal{O}_{ps}$ operator already gives more weight to the high $E_R$ region, so it does not require as large a value of $\delta$ to obtain sufficient events near the observed $248~\mathrm{keV}$ recoil. Therefore, for a fixed $m_{\chi_1}$, the allowed region for $\mathcal{O}_{ps}$ moves towards a smaller $\delta$ compared with $\mathcal{O}_{ss}$. This results in a shift towards the left side for the $TS=2.3$ contour. The shift is not very large because the main $m_{\chi_1}-\delta$ correlation is still controlled by the inelastic kinematics. The difference between the $\mathcal{O}_{ss}$ and $\mathcal{O}_{ps}$ contours comes from the additional $q^2$ dependence of the recoil spectrum. 

The right panel of Fig. \ref{fig:idm_prof} shows the preferred region by the LZ event for the $\mathcal{O}_{pp}$ operator. In this case, the $TS=2.3$ contour is shifted further towards smaller $\delta$ compared with the $\mathcal{O}_{ss}$ and $\mathcal{O}_{ps}$ cases. This larger shift is mainly due to the stronger momentum dependence of the $\mathcal{O}_{pp}$ operator. The recoil spectrum for $\mathcal{O}_{pp}$ contains a $q^4$ dependence, which gives much more weight to larger momentum transfer and therefore to higher recoil energies. As a result, the $\mathcal{O}_{pp}$ spectrum is more strongly weighted towards the high $E_R$ region where the $248~\mathrm{keV}$ event is observed (This is also clearly visible in Fig. \ref{fig:recoilspectra}.). Because of this strong $q^4$ dependence, the $\mathcal{O}_{pp}$ operator already provides a significant contribution at high recoil energies without requiring a large mass splitting. Therefore, for a fixed $m_{\chi_1}$, a smaller value of $\delta$ is sufficient to obtain a recoil spectrum compatible with the observed event. This leads to the more pronounced leftward shift of the $TS=2.3$ contour in the right panel.
The three panels, therefore, show a clear trend. As the momentum dependence becomes stronger, from $\mathcal{O}_{ss}$ to $\mathcal{O}_{ps}$ and finally to $\mathcal{O}_{pp}$, the preferred value of $\delta$ decreases for a fixed $m_{\chi_1}$. Thus, the position of the $TS=2.3$ contour is not determined only by the inelastic kinematics, but also shows the effect of the momentum dependence of the underlying interaction.

We then move to the bDM case. In Fig. \ref{fig:vVSmchi}, we show the allowed region in the $(m_{\chi_1},v)$ plane for the three operators, $\mathcal{O}_{ss}$, $\mathcal{O}_{ps}$, and $\mathcal{O}_{pp}$. The three curves almost overlap, indicating that the kinematic condition for producing the observed $248~\mathrm{keV}$ recoil is nearly independent of the interaction operator. To study the required interaction scale in the bDM scenario, we scan the DM mass together with the momentum of the incoming DM particle. Instead of scanning directly over the velocity, we define
\begin{eqnarray}
{r}_p=\frac{p_{\chi_1}}{p_{\rm th}},
\end{eqnarray}
where $p_{\rm th}$ is the minimum DM momentum required to produce the observed $248~\mathrm{keV}$ recoil for a given $m_{\chi_1}$, {which is coming from Eq. \ref{eq:ERmax}}. We take ${r}_p$ in the range $1\leq {r}_p\leq50$. Thus, ${r}_p=1$ corresponds to the kinematic threshold where $E_R^{\max}=248~\mathrm{keV}$, while larger values of ${r}_p$ correspond to larger incoming DM momentum and hence to $E_R^{\max}>248~\mathrm{keV}$. The momentum is then converted to the DM velocity using the relativistic relation, Eq. \ref{eq:boost}. The scan, therefore, starts exactly from the kinematic boundary and extends to much larger DM momenta, allowing us to determine the interaction scale over the full region where the $248~\mathrm{keV}$ recoil can be produced. Here we fix $\langle\sigma v\rangle_{\chi_2\chi_2\rightarrow\chi_1\chi_1}=10^{-33}{\rm ~cm^2}$.
The resulting values of the profiled interaction scale $\hat{\Lambda}_{i}$ are shown by the color code for the $\mathcal{O}_{ss}$, $\mathcal{O}_{ps}$, and $\mathcal{O}_{pp}$ operators. In all three panels, the boundary of the colored region follows almost the same curve corresponding to $E_R^{\rm max}=248~\mathrm{keV}$. This is because this boundary is determined only by the incoming DM kinematics: $m_{\chi_1}$ and its momentum or velocity, and is independent of the interaction operator. The operators determine the recoil spectrum and the overall event rate, but they cannot change the maximum recoil energy available for a given incoming DM particle. This is why the three operators merge almost completely at the kinematic boundary. The gray region below this boundary corresponds to $E_R^{\rm max}<248~\mathrm{keV}$, where the observed recoil cannot be produced. The horizontal red dashed line shows the maximum allowed DM velocity $v_{\rm max}$, and the red shaded region corresponds to $v<v_{\rm max}$. The $E_R^{\rm max}=248~\mathrm{keV}$ boundary intersects $v_{\rm max}$ at $m_{\chi_1}\simeq78.7~\mathrm{GeV}$. This gives an important mass scale for the bDM scenario. For $m_{\chi_1}<78.7~\mathrm{GeV}$, the velocity required to produce the $248~\mathrm{keV}$ recoil is larger than $v_{\rm max}$, and hence these masses cannot explain the event within the assumed velocity range. For $m_{\chi_1}\gtrsim78.7~\mathrm{GeV}$, the required velocity is below $v_{\rm max}$, and the observed recoil can be produced without requiring any additional boost. Although the kinematic boundary is almost identical for the three operators, the required $\hat{\Lambda}_i$ is clearly different. For $\mathcal{O}_{ss}$, the color scale corresponds to relatively large values of $\hat{\Lambda}_i$, while the required scale becomes much smaller for $\mathcal{O}_{ps}$ and is smallest for $\mathcal{O}_{pp}$. This difference comes from the momentum dependence of the operators. The $\mathcal{O}_{ss}$ operator does not contain an additional power of $q$, whereas $\mathcal{O}_{ps}$ contains a $q^2$ factor and $\mathcal{O}_{pp}$ contains a stronger $q^4$ dependence. Since the observed event is at a high recoil energy, these additional powers of $q$ enhance the contribution from the high momentum transfer region. Consequently, a smaller interaction strength is needed to obtain the same number of events. Since the event rate decreases with increasing $\hat{\Lambda}_i$, this appears as a smaller value of $\hat{\Lambda}_i$ for $\mathcal{O}_{ps}$ and  $\mathcal{O}_{pp}$.

The profiled cutoff scales shown in Figs.~\ref{fig:idm_prof} and ~\ref{fig:vVSmchi} span many orders of
magnitude, and for the pseudoscalar operators they extend below the
momentum transfer characterizing the event.  The condition for a contact description to be valid is that the mediator
be heavy compared with the momentum flowing through it,
$m_\phi\gtrsim q$. This is not the same as requiring $\Lambda\gtrsim q$,
since the effective scale
$
\Lambda^2={m_\phi^2}/{g_\chi g_N}
$
absorbs the couplings of the mediator to the dark and visible sectors as
well as its mass. Taking the mediator as light as the expansion permits,
$m_\phi=q$, and the couplings as large as perturbativity permits,
$g_\chi g_N\le4\pi$, gives the smallest cutoff consistent with the
effective description,
\begin{equation}
\Lambda\;\gtrsim\;\frac{q}{\sqrt{4\pi}}\;\simeq\;69~{\rm MeV}\,,~~
\log_{10}(\Lambda/{\rm GeV})\gtrsim-1.16\,,
\label{eq:eftvalid}
\end{equation}
weaker by a factor $\sqrt{4\pi}$ than the naive requirement
$\Lambda\gtrsim q$. Here $q=\sqrt{2m_NE_R}\simeq246$~MeV is evaluated at
the observed recoil energy. Therefore, we mark the regions of parameter space in Fig. \ref{fig:idm_prof} and Fig. \ref{fig:vVSmchi} with dark magenta-colored hatching, where $\hat{\Lambda}_{i}$ becomes less than 69 MeV. In this region, the EFT description is no longer valid. This limit on $\hat{\Lambda}_i$ can be translated into an upper bound on the DM mass. For a DM velocity of $0.05c$, explaining the LZ event requires $m_{\chi_1}\lesssim 22$ GeV for the $\mathcal{O}_{pp}$ operator in the bDM case.

\section{Conclusion}\label{sec:conclusion}
The single high-energy nuclear recoil event reported by LZ at
$E_R=248$~keV is difficult to accommodate within the standard picture
of elastic scattering of halo dark matter, not because the recoil
energy is unattainable, but because any interaction producing it would
also produce a conspicuous excess at low recoil energy that is not
observed. A viable explanation must therefore supply a spectrum that is
actively suppressed below the observed energy. In this work we have
examined the two kinematically distinct mechanisms capable of doing so: inelastic DM (iDM) and boosted DM (bDM);
treating both within a common model-independent effective operator
framework.

The kinematics itself puts several constraints. Depositing $248$~keV in xenon
requires a momentum transfer $q\simeq246$~MeV and hence a
centre-of-mass momentum $p^\ast\gtrsim123$~MeV. For halo DM
this translates into $\mu_{\chi N}\gtrsim48$~GeV, or
$m_{\chi_1}\gtrsim79$~GeV, and for endothermic scattering it bounds the
mass splitting by $\delta\le385$~keV. For a relativistic flux of light
particles the same condition becomes a requirement on the dark sector
mass ratio, $p_{\chi_1}\gtrsim123$~MeV, which is met for
$\gamma\simeq1.6$ at $m_{\chi_1}=0.1$~GeV. These constraints are
independent of the interaction and our parameter
space is scrutinized against these limits.

Within these bounds we find the following. First, the two scenarios
produce qualitatively different recoil spectra. The inelastic spectra
are concentrated near the observed recoil energy for all three
operators, because the mass splitting forbids low-energy recoils, the fraction of events below $150$~keV ranges from $17.4\%$
to $6.0\%$. 
In contrast, the spectra in the bDM scenario, depend strongly on the
operator. At our benchmark the scalar and pseudoscalar--scalar
interactions place $99\%$ and $92\%$ of their events below $150$~keV,
whereas the pseudoscalar--pseudoscalar interaction places $75\%$ above
it. Thus for a bDM interpretation, the momentum dependence of the
interaction is therefore not a refinement but the essential ingredient,
and the recoil energy of the LZ event by itself prefers
$\mathcal{O}_{pp}$ over the other two operators.

Second, in the inelastic case the preferred mass splitting decreases
monotonically as the momentum dependence of the operator increases.
Additional powers of $q$ shift spectral weight towards higher recoil
energies in the same way that a larger $\delta$ does, so the two
effects are partially interchangeable: an operator that already
concentrates events at high $E_R$ does not require a large splitting to
match the observation. The $1\sigma$ contours accordingly shift towards
smaller $\delta$ from $\mathcal{O}_{ss}$ to $\mathcal{O}_{ps}$ to
$\mathcal{O}_{pp}$. This degeneracy between the operator structure and
the mass splitting is the principal obstacle to inferring $\delta$ from
a single event, and breaking it requires either additional events or a
complementary target.

The prospects for resolving the situation are nonetheless concrete.
Since the expected number of events scales linearly with exposure, the
full LZ dataset and next-generation xenon observatories will either
accumulate a spectrum or exclude these interpretations outright. The
inelastic and boosted scenarios predict different spectral shapes
throughout the high-energy window, and the operators within each
scenario predict different low-energy tails, so even a handful of
additional events would discriminate among the possibilities examined
here. Should the event prove to be a background fluctuation, the
framework developed in this work remains applicable to any future
high-energy recoil candidate.

\section*{Acknowledgment} S.M. acknowledges support from the IIT Goa Startup Grant
[2025/SG/SM/057]. P.K.P. acknowledges the Ministry of Education, Government of India, for providing financial support for his research via the Prime Minister’s Research Fellowship (PMRF) scheme.

%


\appendix

\end{document}